\documentclass{elsart}
\usepackage{latexsym}
\usepackage{graphicx}
\usepackage{subfigure}
\usepackage{amsmath}
\usepackage{color}
\usepackage[numbers,sort&compress]{natbib}

\newcommand{\ie}{i.e.}
\newcommand{\fcc}{f.c.c.}

\newcommand{\Vref}{V_{\text{ref}}}
\newcommand{\eps}{\varepsilon}
\newcommand{\wek}[1]{\boldsymbol{#1}}
\newcommand{\mtr}[1]{\ensuremath{\mathbf{#1}}}

\begin{document}
\begin{frontmatter}

\title{Elastic properties of the degenerate {\fcc} crystal of polydisperse soft dimers at zero temperature}

\author{J.~W.~Narojczyk}\footnote{jwn@ifmpan.poznan.pl} and
\author{K.~W.~Wojciechowski}\footnote{kww@man.poznan.pl
}
\address{Institute of Molecular Physics, Polish Academy of Science,
Smoluchowskiego 17, PL-60-179 Pozna\'n, Poland }

\begin{abstract}
Elastic properties of soft, three-dimensional dimers, interacting through site-site $n$-inverse-power potential, are determined by computer simulations at zero temperature. The degenerate crystal of dimers exhibiting (Gaussian) size distribution of atomic diameters - i.e. size polydispersity - is studied at the molecular number density $1/\sqrt{2}$; the distance between centers of atoms forming dimers is considered as a length unit. It is shown that, at the fixed number density of the dimers, increasing polydispersity causes, typically, an increase of pressure, elastic constants and Poisson's ratio; the latter is positive in most direction. A direction is found, however, in which the size polydispersity causes substantial {\em decrease} of Poisson's ratio, down to {\em negative} values for large $n$. Thus, the system is partially auxetic for large polydispersity and large $n$.
\end{abstract}
\end{frontmatter}

\section{Introduction}

Among the characteristics of real materials, elastic properties are the ones that are crucial from the point of view of various applications. Recently, increasing interest is observed in the area of materials with unusual elastic properties - an example are materials with negative Poisson's
ratio~\cite{Lak1987Scien_1,Eva1991Endeavour,Lak1993AdvMat,BauShaZak1998Nature,%
EvaAld2000EngSciEdJou,Bau2003Nature,RemScaKww2009PSSb}. When stretched (or resp. compressed) such materials increase (resp. decrease) their size not only in the direction of the stretch (resp. compression) but also in one or more directions transverse to it~\cite{RemScaKww2009PSSb}. Materials exhibiting such counterintuitive behavior are known as anti-rubber~\cite{stronaInternetowaLakesa}, dilational materials~\cite{Mil92JMPS} or {\em auxetics}~\cite{Eva1991Endeavour}. In this paper, the last name will be used for its brevity.

As for isotropic systems the Poisson's ratio is direction independent, any isotropic material can be either auxetic or not. For anisotropic systems, which are the subject of the present paper, the situation is more complex. Apart from the alternative that the Poisson's ratio is in all directions negative or not, a third possibility exists: in some directions it is negative and in other - it is not. In the literature related to the field, various names have been proposed to describe different situations~\cite{BauShaZak1998Nature,Kww2005CMST,TinBar2005JAM,BraHeyKWW2009PSSb}. Here we will adopt the scheme introduced in Ref.~\cite{BraHeyKWW2009PSSb}, in which - as in the isotropic case - the names auxetics and nonauxetics correspond, respectively, to negative or nonnegative Poisson's ratio in all directions. Materials with Poisson's ratios negative in some directions and nonnegative in the other ones will be referred to as {\em partially} auxetic materials or just {\em partial auxetics}.

The very existence of auxetics and partial auxetics encourages one to study influence of various micro-level mechanisms on the elastic properties of condensed matter~\cite{Alm1985JEL,KWW1987MP,BatRot1988IJES,KWW1989PLA,KWWBra1989PRA,%
EvaNkaHut1991Nature,RotBerBat1991Nature,Wei92JCP,BoaSeiShi1993PRE,Kww1995MPR,%
PraLak1997IJMS,WeiEdw98PRE,BauETAL2000Science,GriEva2000JMSL,IshIwa2000JPSJ,%
KimKabKog2000PRL,SmiGriEva2000AM,BowCatThoTra2001PRL,VasDmiIshShi2002PRE,%
KWW2003JouPhyA,XinRad2003PRL,KwwKvtMko2003PhysRevE,%
Pik2004PRL,GriETAL2007PSSb,AttGri2008PSSb,GriGatFar2008PSSb,HalETAL2008Sci,%
RecStiTor2008PRL,LakKWW2009PSSb,AttManGri2009PSSb,AttManGatGri2009PSSb,PikETAL2009PSSb}.
An important problem in this context is the influence of disorder on the Poisson's ratio. For some two-dimensional (2D) models which are elastically isotropic (at least for small deformations) and for some anisotropic 2D models, it has been found that various kinds of disorder (such as vacancies, interstitials, aperiodicity or polydispersity) introduced in crystalline structures, increase Poisson's ratio~\cite{KwwKvt1999CPC,KvtKww2005PhysStatSolB,JwnKww2007PSSB,JwnKww2006JNCS,
JwnKww2008PSSB_2,JwnAldImr2008JNCS}. Studies of some three-dimensional~(3D) molecular models~\cite{MkoKww2005PhysStatSolB} revealed similar effect caused by vacancies, what might suggest a general conclusion that (at least in simple molecular systems) disorder weakens or eliminates auxeticity.

Recently, however, it has been found that in one of the simplest 3D systems - soft spheres, forming the {\fcc} phase, there exists a direction in which one of the form of disorder (polydispersity) {\em decreases} Poisson's ratio, which reaches negative values down to -1~\cite{JwnKww2008PSSB_1}. This observation might find practical applications, e.g. in production of partial auxetics of latex meso- or macroscopic spheres for which size polydispersity is easy to control. Unfortunately, the mentioned lowest Poisson's ratios were observed only for the softest spheres with $n=6$, i.e. when the {\fcc} structure may be unstable with respect to the b.c.c. structure~\cite{HooYouGro1971JCP}.

Taking into account that elastic properties of many-body systems strongly depend on details of the intermolecular interaction potential and, in particular, on the molecular shape, it is interesting to study other 3D systems searching for stable structures for which increasing polydispersity can lead to more negative Poisson's ratios in some directions. As it is meaningful to start with the simplest molecular shapes, the natural candidate for a first 3D molecule to investigate is a polydisperse dimer.

The aim of this study is to determine the influence of disorder introduced to a solid of soft dimers by the size dispersion of their atoms, on the elastic properties of the model. The study concerns the zero temperature, $T=0$, and the number density $N/V=1/\sqrt{2}$, where $V$ is the volume and $N$ is the number of dimers. At this density atoms of dimers can be arranged in a perfect {\fcc} lattice
of lattice constant equal to the distance of atomic centers in the dimer. The dimers form a degenerate crystalline (DC) phase - an analogue of that observed in 2D~\cite{KwwFreAcb1991PhysRevLet}. The phase is characterized by non-periodic positions and orientations of molecules, whereas the atoms forming molecules are arranged in a perfect {\fcc} lattice (when the atoms are identical) or (for slightly different sizes of atoms) in a lattice close to the {\fcc} one. The DC phase is known to be thermodynamically stable and effectively cubic from the point of view of elastic constants for hard dimers~\cite{Mko2005phd} and is expected to preserve those properties for soft dimers with short ranged interactions which are studied in the present paper.

The soft dimer system can be seen as a rough model of some man-made \linebreak nano-, meso- or macro-particle systems, e.g. systems of latex particles~\cite{HooLanRom2004PRE,ManPin2004MRS}. Recently, an increasing interest is observed in systems of soft particles~\cite{HeyBra2007PCCP}, especially in the context of physics of colloids. This is due to the fact that the latter have
a broad range of possible applications. Systems of soft polydisperse particles seem to be more realistic models of colloids than those built of identical particles. Polydispersity is also important for nanoparticle crystals~\cite{KalFialPas2006Sci,KalGrz2007NanoL,PinKalKow2007JPC},
where the speed of crystallization and resulting crystal quality is strongly related to size dispersion of particles.

Studies of elastic properties of models of polydisperse systems can be thought of as a first step on a path to better understanding mechanisms and phenomena that govern the elastic properties of real systems. Such a knowledge is not only important from the point of view of fundamental research but also useful from the material engineering point of view - it can help to engineer materials with desired elastic properties.

\section{Description of the studied model}

The model studied in this work consists of $N$ diatomic molecules (dimers) in periodic boundary conditions. The dimers are rigid {\ie} the distance between centers of the atoms forming
each one is constant and equal to $\sigma$ (which constitutes a unit of length) for each dimer.
Initially the atoms form the perfect {\fcc} lattice and the molecules form a degenerate crystalline (DC) phase. The atoms interact {\em via} purely repulsive potential of the form:
\begin{equation}
u_{ij}(d_{ij},r_{ij})=u_0\left(\frac{d_{ij}}{r_{ij}}\right)^n\ ,
\label{EQip}
\end{equation}
where $d_{ij}=(d_i+d_j)/2$, $d_i$, $d_j$  are the diameters of the interacting spheres (atoms), $r_{ij}$ is the distance between their centers and $u_0$ is the energy unit. The exponent $n$ is further referred to as the \textit{hardness parameter} and its inverse, $1/n$, as the \textit{softness parameter}. (These names come from the observation that when $n\rightarrow\infty$ the interaction potential described by Eq.~(\ref{EQip}) tends to the hard-body potential, which is widely used in computer simulations and condensed matter theory~\cite{KwwKvtMko2003PhysRevE}.)
Atomic interactions are assumed to be short-range ones, {\ie} only the nearest neighbors (shearing a face of their Voronoi polyhedra) interact. Since the molecules are considered as rigid, interactions between atoms forming a molecule are neglected.
\begin{figure}[b]
 \includegraphics[width=0.42\textwidth]{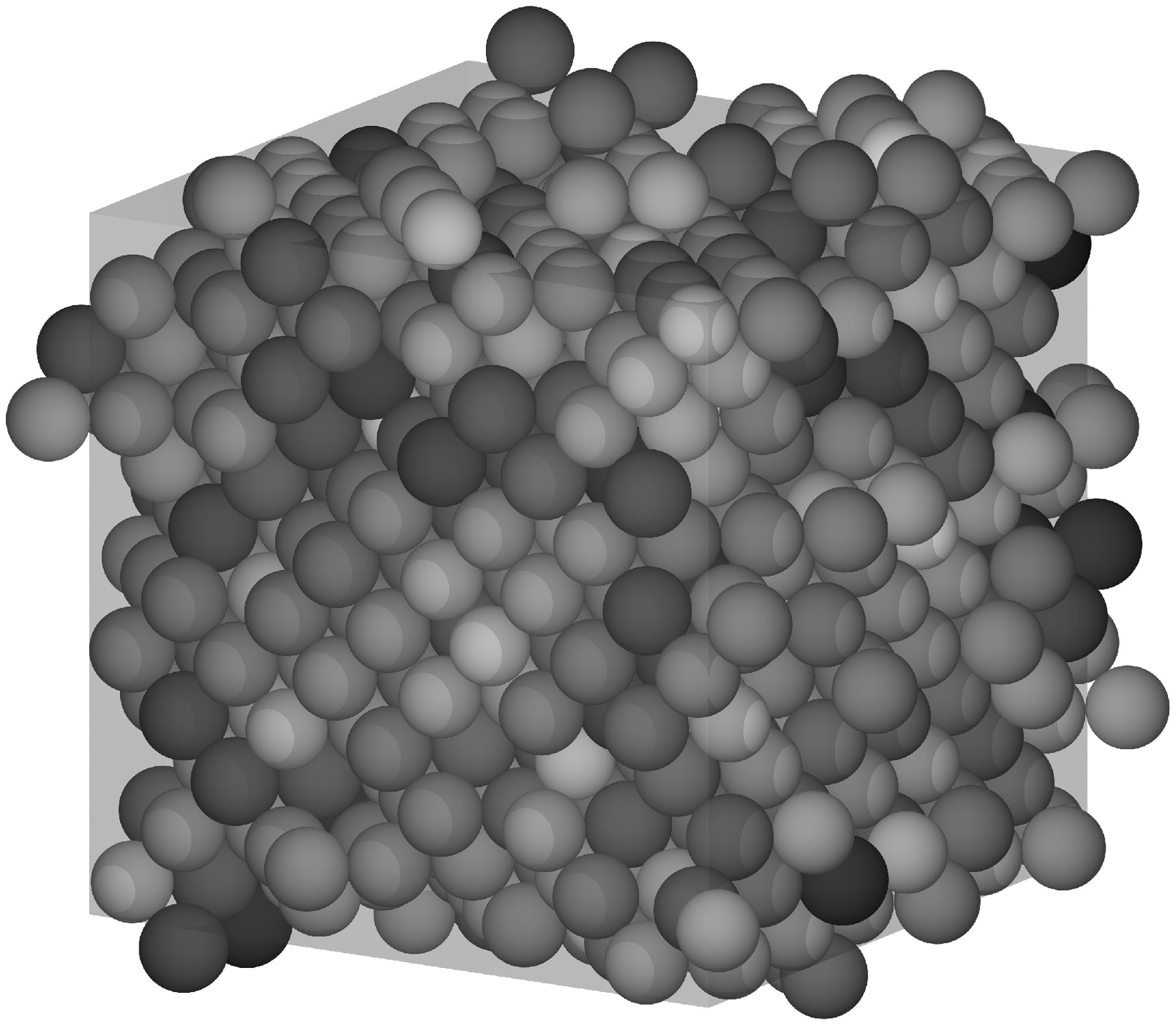}~a)
 \hfill
 \includegraphics[width=0.42\textwidth]{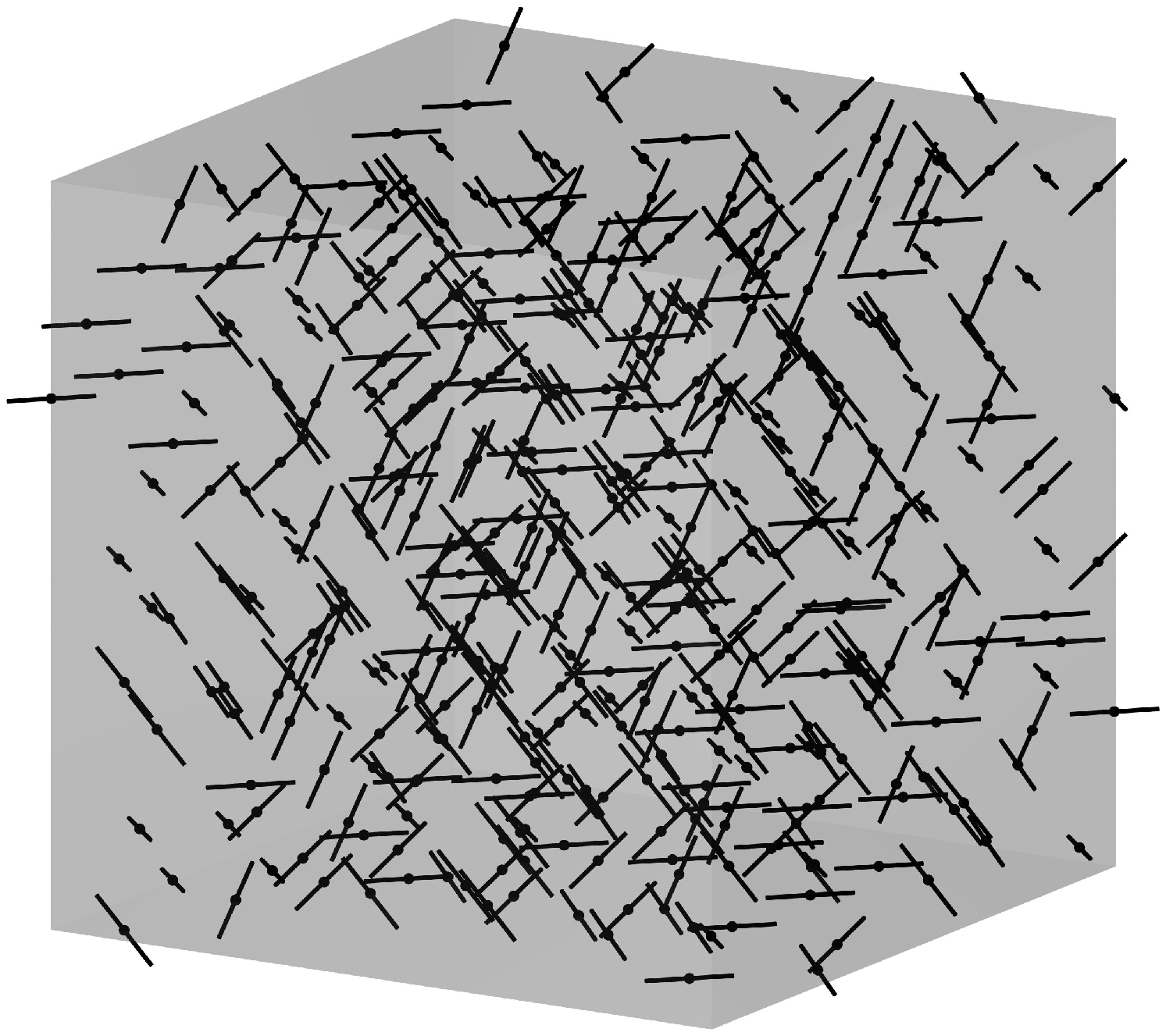}~b)
 \caption{Example of studied structures with atomic size polydispersity.
    The structure shown consists of $432$ molecules and has the polydispersity
    parameter $\delta$ equal to~$3\%$. In~(a) the
    atoms are drawn in the shades of gray, indicating their size
    relative to other atoms. The ones drawn in white are the smallest
    and those drawn in black are the largest. The gray cube indicates
    the periodic box.
    In (b)~connections between atoms in the same structure as in~(a) are shown.
}
 \label{FG_graphic_structures}
\end{figure}

\subsection{Atomic size polydispersity}
In the studied system, atomic diameters were generated randomly according to the
Gauss distribution function with a given standard deviation $\delta$
\begin{equation}
\delta=\frac{1}{\sigma} \sqrt{\left< d^2\right>-\left< d\right>^2}\ .
\label{EQdelta}
\end{equation}
In above equation, $d$ is the atomic diameter and $\sigma=\left<d\right>$.
Further, in this work, $\delta$ is treated as the measure
of disorder in the system and is referred to as the {\em polydispersity
parameter}. Generated values of atomic diameters were randomly distributed
among atoms in the structure. In the Fig.~\ref{FG_graphic_structures} examples
of studied structures are presented. It can be seen that even in the presence
of atomic size polydispersity atoms form nearly perfect {\fcc} lattice.

\section{Basic formulae}
\label{sec:basic} In the following, it will be assumed that the studied
system of soft dimers exhibits effective regular symmetry. For the
latter symmetry, when the external pressure $p$ is zero, elastic properties can be
described by just three elastic constants $C_{xxxx}$, $C_{xxyy}$ and
$C_{xyxy}$~\cite{LanLif1986BOOK:Per}. However, when $p\neq 0$, it is more
convenient~\cite{Kww1995MPR,KwwKvt1999CPC}
to use other elastic constants, $B_{ijkl}$, obtained by expanding the
free enthalpy (Gibbs free energy), $G$, with respect to the components
of the Lagrange strain tensor, $\eps$:
\begin{equation}
 \Delta G/\Vref = \Delta(F+pV)/\Vref=\frac{1}{2} \sum_{ijkl} B_{ijkl} \eps_{ij}\eps_{kl}\ .
\label{EQ_FreeEnthalpyGeneral}
\end{equation}
$F$ in the above equation is the (Helmholtz) free energy and $\Vref$ is the volume of the deformation-free (reference) system. The elastic constants $B_{ijkl}$ are simply related with the constants $C_{ijkl}$~\cite{Wal1972:Wil,KvtKww2005JouChemPhys}
\begin{equation}
B_{ijkl}=C_{ijkl} -p\left(\delta_{ik}\delta_{jl} +\delta_{il}\delta_{jk} -\delta_{ij}\delta_{kl}\right)\ ,
\label{EQ_BtoC}
\end{equation}
where $\delta_{ij}$ is the Kronecker delta, equal to 1 for $i=j$ and zero otherwise.

In the case of cubic symmetry, considered in this work, only the following elastic constants
are of interest (the Voigt notation is used below~\cite{Wal1972:Wil}):
\begin{eqnarray}
B_{11}&=&C_{11}-p\ ,\label{EQ_BC11}\\
B_{12}&=&C_{12}+p\ ,\\
B_{66}&=&C_{66}-p\ ,\label{EQ_BC66}
\end{eqnarray}
and the free enthalpy expansion (\ref{EQ_FreeEnthalpyGeneral}) takes the form:
\begin{equation}
\begin{split}
\Delta G/\Vref&= \frac{1}{2}B_{11}
 \left(\eps_{xx}^2+\eps_{yy}^2+\eps_{zz}^2\right)\\&+
 B_{12}\left(\eps_{xx}\eps_{yy}+\eps_{yy}\eps_{zz}+
  \eps_{zz}\eps_{xx}\right)+ 2B_{66}\left(\eps_{xy}^2+
  \eps_{yz}^2+\eps_{zx}^2\right)\ .
\end{split}
\end{equation}

For cubic systems, instead of constants defined by Eq.~(\ref{EQ_BC11})-(\ref{EQ_BC66}),
the below elastic moduli are also often used:
\begin{eqnarray}
B&=&\frac{1}{3}\left(B_{11}+2B_{12}\right)\ ,\\
\mu_1&=&B_{66}\ , \\
\mu_2&=&\frac{1}{2}\left(B_{11}-B_{12}\right) \ ,
\end{eqnarray}
where $B$ is the {\em bulk modulus}, and the $\mu_i$ are the {\em shear moduli}. In the following, both sets of the quantities will be used while presenting and discussing the results.

The Poisson's ratio is defined as the negative ratio of transverse to longitudinal strain when only the longitudinal component of the stress tensor is infinitesimally changed. For anisotropic systems it depends, in general, on both the longitudinal and the transverse direction. For cubic symmetry,
the Poisson's ratio along two high-symmetry longitudinal directions ($[100]$ and $[111]$) does not depend on the choice of the transverse direction~\cite{Kww2005CMST} and can be expressed by the elastic constants, $B_{ij}$, in the form~\cite{MkoKww2006JNCS}:
\begin{eqnarray}
\nu_{\left[ 100 \right] } &=& \frac{B_{12}}{B_{11}+B_{12}}
=\frac{1}{2}\frac{3B - 2\mu_2}{3B + \mu_2} \ ,\label{EQnu100}
\\
\nu_{\left[ 111 \right] } &=& \frac{B_{11} + 2 B_{12} - 2 B_{66}}
{2\left(B_{11} + 2 B_{12} + B_{66}\right)}
=\frac{1}{2}\frac{3B - 2\mu_1}{3B + \mu_1}\ .
\label{EQnu111}
\end{eqnarray}
In this paper, the Poisson's ratio along the longitudinal direction $[110]$ and two transverse directions $[1\bar{1}0]$ and $[001]$ was also computed:
\begin{equation}
\begin{split}
    \nu_{\left[ 110 \right]\left[ 1\bar{1}0 \right] }
    &= \frac{B_{11}^2 + B_{11}B_{12} - 2 B_{12}^2 - 2 B_{11}B_{66}}
    {B_{11}^2 + B_{11}B_{12} - 2 B_{12}^2 + 2 B_{11}B_{66}}\\
    &=\frac{3B(3\mu_2 -\mu_1) - 4\mu_1\mu_2}
     {3B\left( 3\mu_2 + \mu_1 \right) + 4\mu_1\mu_2}\ ,
\end{split}
\label{EQnu110110}
\end{equation}
\begin{equation}
\begin{split}
 \nu_{\left[ 110 \right]\left[ 001 \right] }
&= \frac{4 B_{12} B_{66}}
{B_{11}^2 + B_{11}B_{12} - 2 B_{12}^2 + 2 B_{11}B_{66}}\\
&=\frac{6B\mu_1 - 4\mu_1\mu_2}{
 3B\left( 3\mu_2 + \mu_1\right) + 4\mu_1\mu_2}\ .
\end{split}
\label{EQnu110001}
\end{equation}

\section{Numerical determination of the elastic properties}
In order to obtain the elastic constants, a sample was considered which, at
equilibrium, is  described by three vectors $\wek{a}_0$, $\wek{b}_0$
and~$\wek{c}_0$ parallel to the unit cell edges of the
{\fcc} lattice. Components of those vectors form columns of a {\em reference
box matrix}~\cite{ParRah1981JAP,ParRah1982JouChemPhys} $\mtr{H}$.
Applying a uniform deformation that transforms the sample into a
parallelepiped described by the box matrix
$\mtr{h}$ one can write the Lagrange strain
tensor as~\cite{ParRah1981JAP,ParRah1982JouChemPhys}:
\begin{equation}
\mtr{\eps}=\left(\mtr{H^\prime}^{-1}\ \mtr{h^\prime}\ \mtr{h} \ \mtr{H}^{-1} - \mtr{I}\right)/2\ ,
\end{equation}
where $\mtr{h'}$ is the transposed matrix $\mtr{h}$ and $\mtr{I}$ is the unit
matrix. Differentiating the free energy $F$ with respect to the strain tensor
components,
one obtains all the elastic constants and pressure. Is is worth to note,
that in the case of zero temperature ($T=0^\circ$K), the free energy is just
equal to the potential energy of the system.

After determining equilibrium positions and orientations of all the molecules in
the reference state for a given $\delta$, the following deformations were
applied:
\begin{equation}
h_{\alpha\beta}=\left(1+\xi_{\alpha}\right)
H_{\alpha\beta}\delta_{\alpha\beta}\ ,
\label{EQ_deform1}
\end{equation}
for which new equilibrium positions and orientations were determined.
In~(\ref{EQ_deform1}) $\alpha,\beta$ indicate directions, and $\xi_\alpha$ is a small real
number. The relations below define the elastic constants and pressure:
\begin{eqnarray}
\left. \frac{\partial F}{\partial \xi_\alpha}\right|_{\xi_\alpha=0}&=&-p\ , \\
\left. \frac{\partial^2 F}{\partial \xi_\alpha^2}\right|_{\xi_\alpha=0}&=&
        B_{11}\ , \label{EQ_dF2D1}\\
\left. \frac{\partial^2 F}{\partial \xi_\alpha \partial \xi_\beta}
        \right|_{\xi_\alpha=\xi_\beta=0,\ \alpha
        \ne \beta}&=& B_{12}-p\ . \label{EQ_dF2D2}
\end{eqnarray}

The $B_{66}$ constant was obtained by applying the deformation of the form:
\begin{eqnarray}
h_{\alpha\alpha}&=&H_{\alpha\alpha}\ ,\notag\\
h_{\alpha\beta}&=&H_{\alpha\beta}+\zeta_{\alpha\beta} H_{\alpha\alpha}\ ,
\label{EQ_deform2}
\end{eqnarray}
for which new equilibrium positions and orientations were determined.
In~(\ref{EQ_deform2}) $\zeta_{\alpha\beta}=\zeta_{\beta\alpha}$ is a small real number. For
$H_{xx}=H_{yy}=H_{zz}$ this leads to:
\begin{equation}
\left.\frac{\partial^2 F}{\partial\zeta_{\alpha\beta}^2}
        \right|_{\zeta_{\alpha\beta}=0}=
        6\left( 2B_{66}+p\right)\ .
\end{equation}

\section{Results}
\begin{figure}[tp]
 \includegraphics[width=0.4\textwidth]{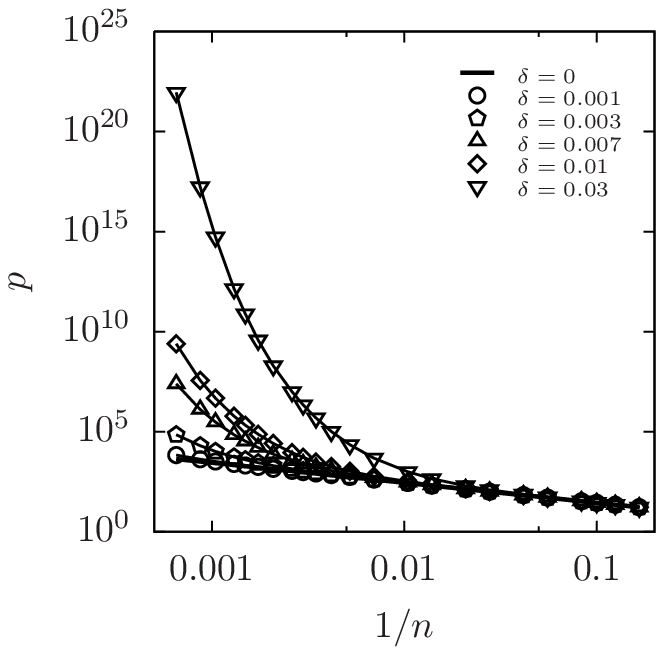}~a)
 \hfill
 \includegraphics[width=0.4\textwidth]{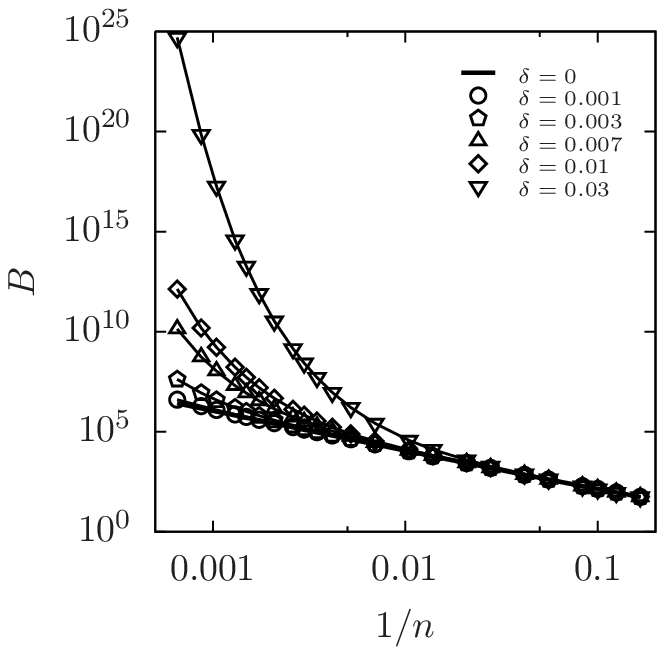}~b)
 \caption{The values of (a)~pressure and (b)~bulk modulus B for polydisperse
 dimers, plotted against the softness parameter ($1/n$). The thick solid line represents
 values for the system with $\delta=0$.}
 \label{FG_6_pressure_and_bulk}
\end{figure}

In the Figs.~\ref{FG_6_pressure_and_bulk}~(a) and~\ref{FG_6_pressure_and_bulk}~(b) values of the pressure and the bulk modulus are shown, respectively. Is is clearly visible that increasing polydispersity and increasing $n$ leads to increase of those quantities as it was earlier observed in other systems~\cite{JwnKww2006JNCS,JwnKww2008PSSB_2,JwnKww2007PSSB,JwnAldImr2008JNCS,JwnKww2008PSSB_1}.

The observed behavior can be qualitatively understood by noticing that dense packing of identical molecules in the hard body limit requires less space than dense packing of polydisperse molecules of the same average size~\cite{KvtKww2005PhysStatSolB}. As all the considered systems are studied at the same volume, their densities relative to close packing grow with increasing polydispersity. Increase of the relative density implies an increase of the pressure and the bulk modulus. Obviously, the latter quantities show the largest increase for the largest $n$, when the interaction potential is the steepest one.

\begin{figure}[htp]
 \includegraphics[width=0.28\textwidth]{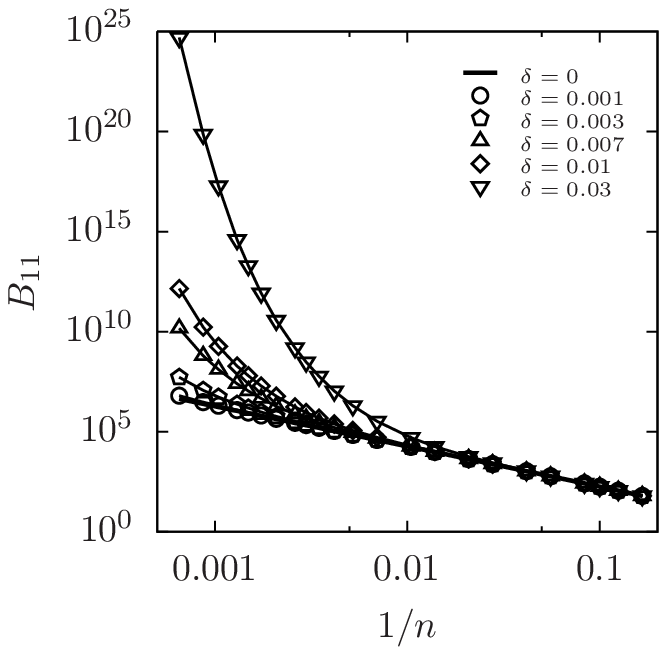}~a)
 \hfill
 \includegraphics[width=0.28\textwidth]{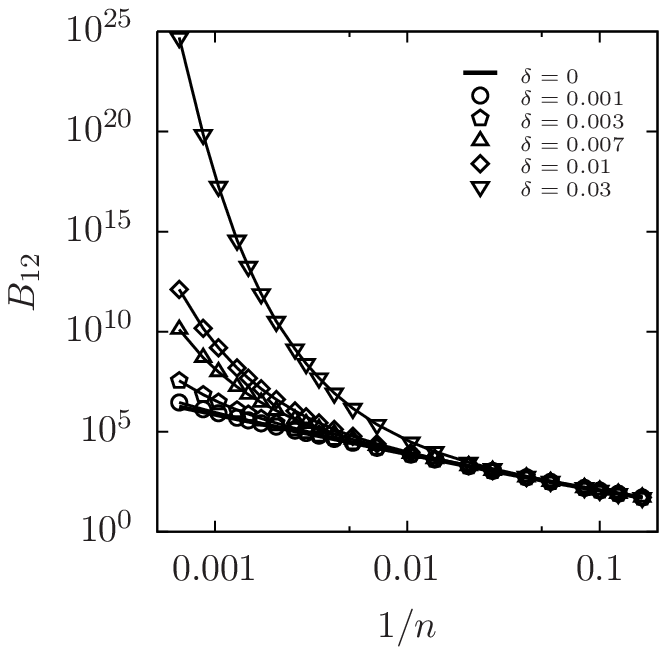}~b)
 \hfill
 \includegraphics[width=0.28\textwidth]{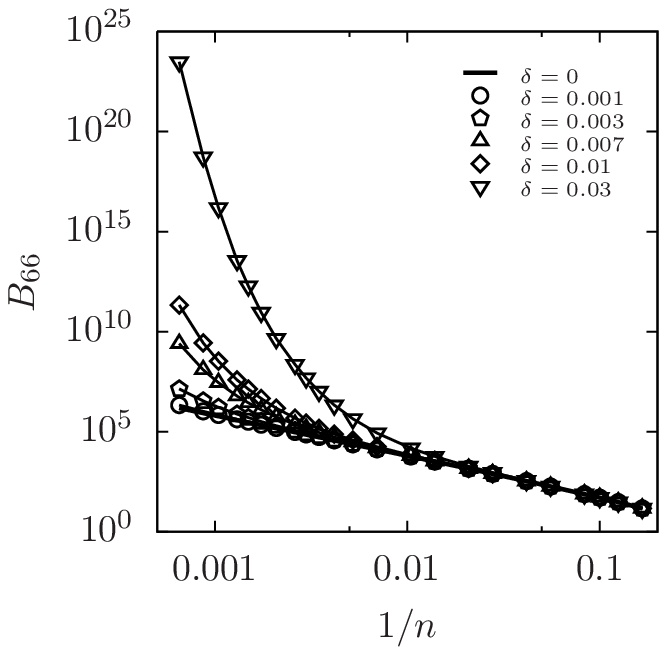}~c)
 \caption{The values of elastic constants (a)~$B_{11}$, (b)~$B_{12}$ and (c)~$B_{66}$.
    plotted against the softness parameter $(1/n)$. The thick solid line represents
    values for the system with $\delta=0$.}
 \label{FG_6_elastic_constants}
\end{figure}

In Fig.~\ref{FG_6_elastic_constants} the elastic constants $B_{ij}$ are shown. It can be seen that all elastic constants also increase with increasing polydispersity and increasing vaule of $n$. Presented behavior of elastic constants is also similar to that observed for other 2D and 3D
systems~\cite{JwnKww2006JNCS,JwnKww2008PSSB_2,JwnKww2007PSSB,JwnAldImr2008JNCS,JwnKww2008PSSB_1} and can be qualitatively explained by the close packing argument used above.

\begin{figure}[htp]
 \includegraphics[width=0.46\textwidth]{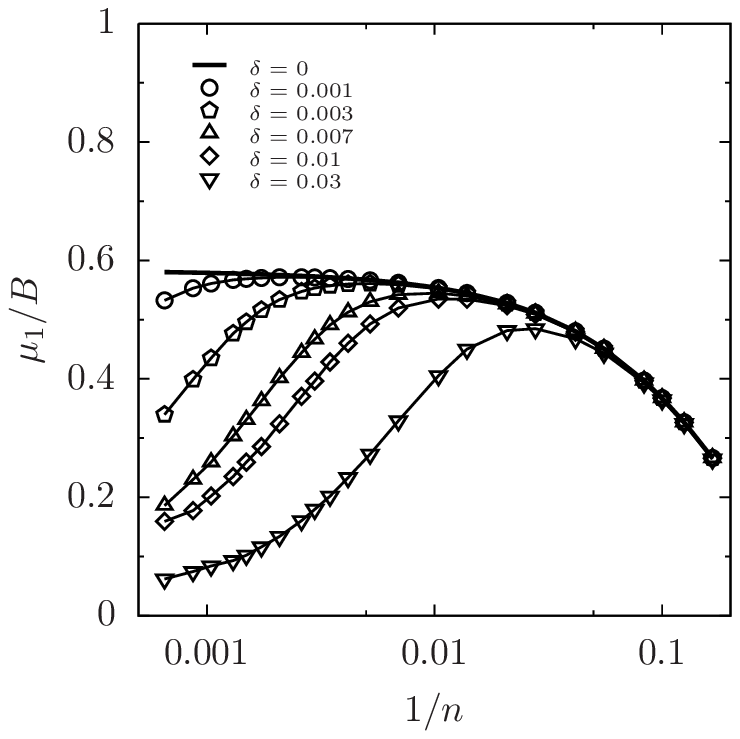}~a)
 \hfill
 \includegraphics[width=0.46\textwidth]{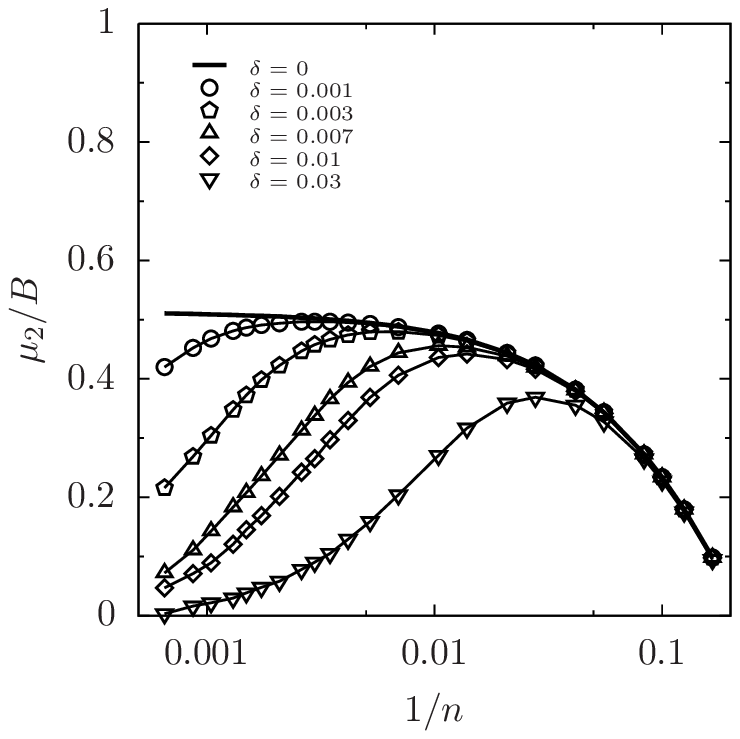}~b)
 \caption{The values of $\mu/B$ ratio for (a)~$\mu_1$ and (b)~$\mu_2$,
 plotted against the softness parameter ($1/n$). Thick solid line represents
 values measured for the system with $\delta=0$.}
 \label{FG_6_shear_bulk_ratio}
\end{figure}

In Fig.~\ref{FG_6_shear_bulk_ratio} plots of $\mu_i/B$ ratios are presented. As in the case of
earlier studies~\cite{JwnKww2006JNCS,JwnKww2007PSSB,JwnAldImr2008JNCS,JwnKww2008PSSB_1,JwnKww2008PSSB_2} for both 2D and 3D, this ratios seem to tend to $0$ for {\em any} nonzero polydispersity, in the hard interactions limit ($n\rightarrow\infty$).

\begin{figure}[htp]
 \includegraphics[width=0.46\textwidth]{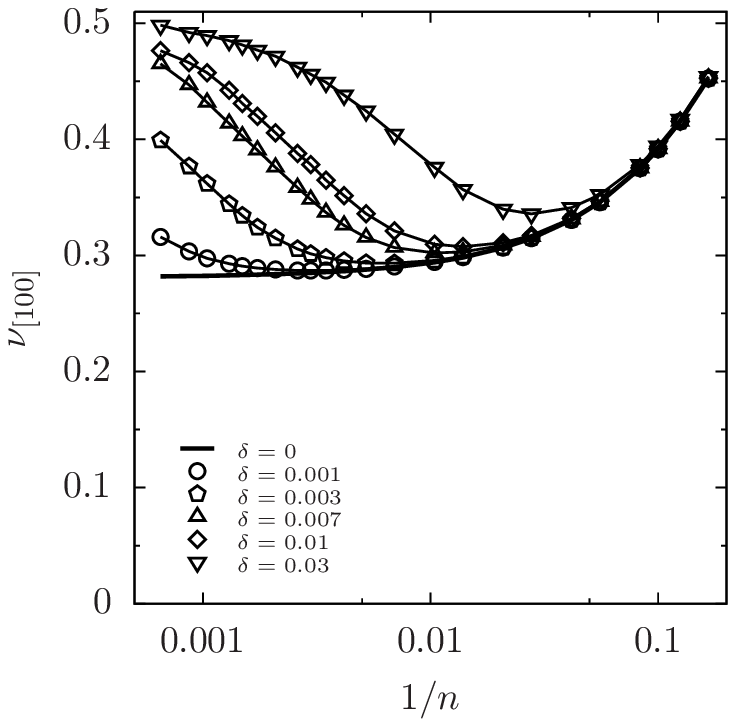}~a)
  \hfill
 \includegraphics[width=0.46\textwidth]{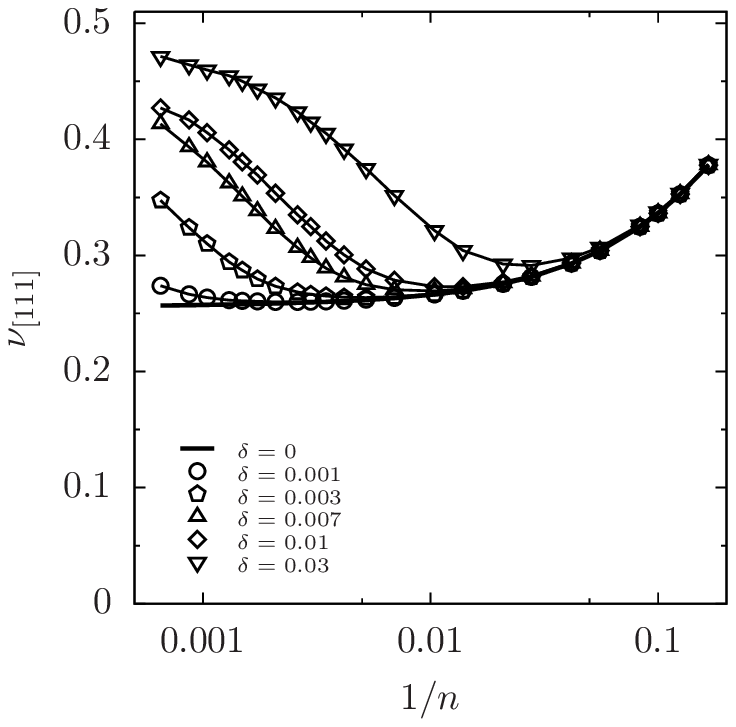}~b)
 \includegraphics[width=0.46\textwidth]{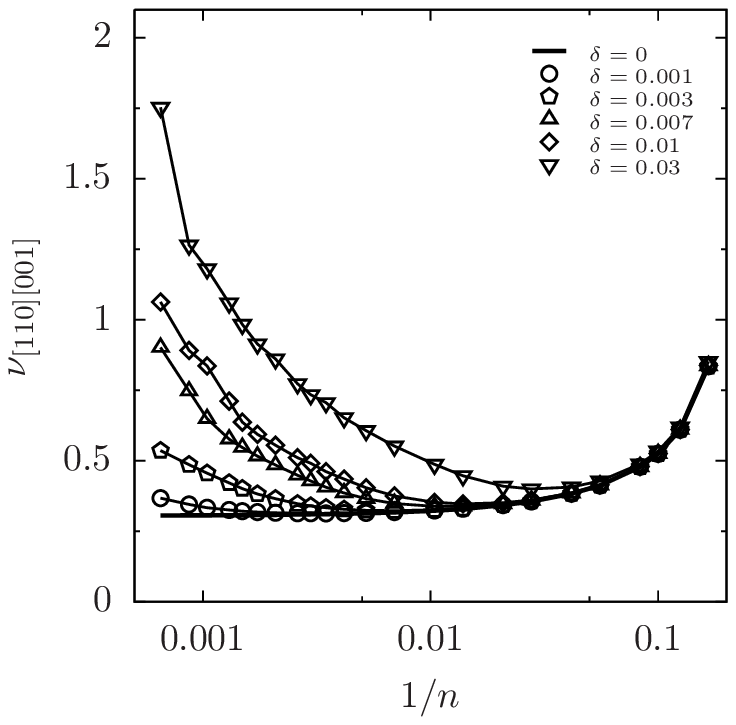}~c)
  \hfill
 \includegraphics[width=0.46\textwidth]{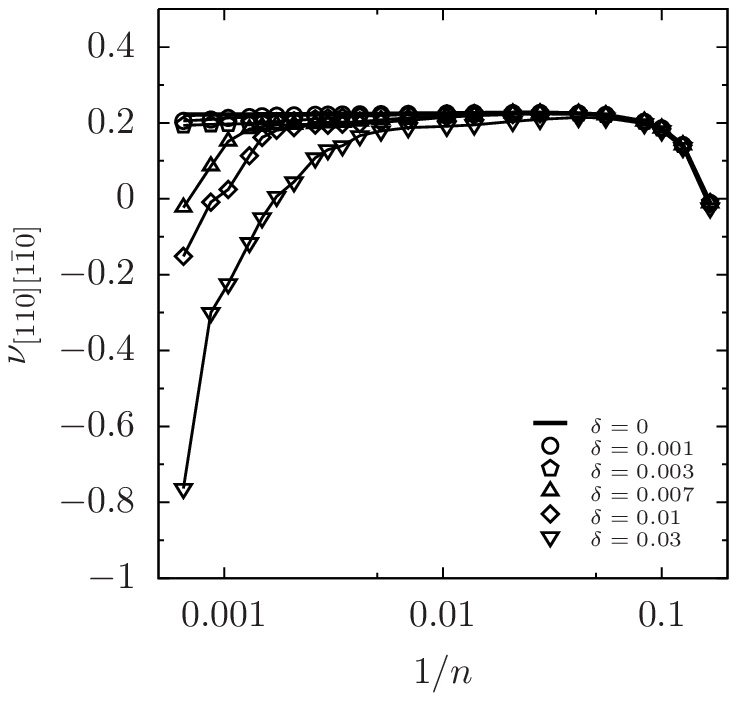}~d)
 \caption{Poisson's ratio plotted against the softness parameter ($1/n$).
 Plots (a),~(b)~show the Poisson's ratio measured perpendicularly to
 the direction of the high-symmetry axis (a)~$[100]$ and (b)~$[111]$.
 Plots (c),~(d)~show Poisson's ratio  perpendicularly to the direction $[110]$
 in the direction (c)~$[001]$ and (d)~$[1\bar{1}0]$, respectively.}
 \label{FG_6_Poissons_ratio}
\end{figure}

According to the formulae (\ref{EQnu100}) and (\ref{EQnu111}) the mentioned above asymptotics of $\mu_i/B$ implies that the system behaves like rubber ($\nu_{rubber}=0.5$) when a small stress is applied along the directions [100] and [111] in the limit $n\to\infty$. This is shown in Fig.~\ref{FG_6_Poissons_ratio}, where the plots of Poisson's ratio are presented. It can be seen there that, typically, presence of any nonzero polydispersity causes an increase of the Poisson's ratio in most of directions corresponding to the high symmetry axes. It is worth to notice that in the cubic system studied, the Poisson's ratio $\nu_{[110][001]}$ {\em exceeds} the maximum value 1/2 allowed for isotropic materials. One should also stress that a pair of exceptional directions (see Fig.~\ref{FG_6_Poissons_ratio}(d)) exists in the system for which the Poisson's ratio {\em decreases} with increasing polydispersity. This occurs when the direction of the deformation is $[110]$ and the response of the system is measured in the direction $[1\bar{1}0]$. For this pair of directions, the Poisson's ratio reaches {\em negative} values both for small and large $n$ limits!
The decrease for large $n$ is, however, much more significant. This new behavior of Poisson's ratio is opposite to what was discovered in the soft polydisperse spheres system~\cite{JwnKww2008PSSB_1}, where the most negative values of Poisson's ratio were observed for small $n$ only.

The observed dependence of $\nu_{[110][1\bar{1}0]}$, surprising in the context of results obtained for various 2D isotropic systems including the DC phase of 2D dimers~\cite{JwnKww2008PSSB_2}, is of interest from the point of view of production of partial auxetics. As shown in Fig.~\ref{FG_6_Poissons_ratio}(d), solid structure of dimers obtained by `gluing' together pairs of polydisperse soft spheres arranged in the {\fcc} lattice will show the Poisson's ratio as low as $\nu_{[110][1\bar{1}0]}= -0.8$.

\section{Conclusions}
The elastic properties of polydisperse soft dimer system in three dimensions were studied by computer simulations in the static (zero temperature) case. Disorder was introduced to the DC {\fcc} structure in the form of atomic size dispersion with polydispersity parameter $\delta$.
The influence of polydispersity on elastic properties of 3D DC dimers system was determined. It was found that increasing polydispersity and hardness of the interaction potential causes a significant increase of elastic constants and pressure.

The simulations have shown that Poisson's ratio of the dimer system measured in $[100]$ and $[111]$ directions increases to its maximum positive value, $1/2$, in the presence of any nonzero polydispersity and sufficiently hard interaction potential. An increase of Poisson's ratio was observed also in other directions. Such a behavior was earlier observed in the case of
2D~\cite{JwnKww2006JNCS,JwnKww2007PSSB,JwnAldImr2008JNCS,JwnKww2008PSSB_2} and 3D~\cite{JwnKww2008PSSB_1} systems and seems to be a typical result.

It is worth to stress, however, that for the longitudinal direction $[110]$ and transverse direction $[1\bar{1}0]$ the Poisson's ratio {\em decreases} reaching {\em negative} values along with increasing polydispersity value both for large and small $n$. This unusual phenomenon is of interest from the point of view of manufacturing stable auxetics of cubic symmetry.

The questions whether the effect of auxeticity enhancement by polydispersity is characteristic for the {\fcc} lattice only or if it can be observed for elastic symmetries other than the {\fcc} one, e.g. for packing of spheres in the h.c.p. lattice or stacking fault lattices, will be a subject of a separate paper. Studies of elastic properties of other molecular shapes are in progres.

\section*{Acknowledgements}
This work was partially supported by the grants NN202 070 32/1512 and NN202 261438 MNiSzW. We are grateful to Dr. Miko{\l}aj Kowalik for preparing the samples of the DC phase of the dimers, used in the simulations. Most of the simulations were carried out at the Pozna\'n  Supercomputing
and Networking Center (PCSS).


\newpage

\end{document}